\documentclass[pra,showpacs,twocolumn,epsfig]{revtex4}
\usepackage{amsfonts}
\usepackage{amsmath}
\usepackage{amssymb}
\usepackage{graphicx}

\begin{document}
\draft
\title{Improvement of conversion efficiency of atom-molecule Bose-Einstein
condensate}
\author{Guang-Ri Jin \cite{email} and Sang Wook Kim \cite{email1}}

\address{Department of Physics Education and Department of Physics, Pusan National University, Busan 609-735, Korea}
\date{\today}

\begin{abstract}
We investigate the stimulated Raman adiabatic passage in two-color photoassociation for a atom-molecule Bose-Einstein condensate. By applying
two time-varying Guassian laser pulses that fulfill generalized two-photon resonance condition, we obtain highly efficient atom-molecule
conversion. The efficiency depends on the free-bound detuning and the delay time between the two pulses. By adjusting the parameters optimally,
we achieve $92\%$ conversion efficiency.
\end{abstract}


\maketitle

\section{Introduction}

Possible realization of molecular Bose-Einstein condensate (BEC) from atomic condensate has attracted wide attentions
\cite{Wynar,drum,Java99,hein,timm,Donley}. A pair of atoms in open channel can be coupled to a weakly bound molecule by using either
photoassociation process \cite{drum,Java99,hein} or Feshbach resonance method \cite{timm,Donley}. Such a coherent conversion of two chemically
different species is of fundamental physical interests, and can be regarded as matter-wave analog of second harmonic generation. It was shown
that a direct conversion via one-color free-bound photoassociation (Feshbach resonance) creates the molecules in highly excited electronic
(vibrational) levels. Moreover, statistical property of the molecules exhibits sub-Poissonian only at initial stage \cite{Meystre}, and it will
be transformed into super-Poissonian due to inherent nonlinearity of the BEC \cite{GRJIN}.

To overcome all these problems, the second laser field is employed
to couple the excited molecules to long-live ground-state molecules.
Two-color free-bound-bound mechanism based on the stimulated Raman
adiabatic passage (the STIRAP) was proposed as a promising route to
achieve the high efficiency of atom-molecule conversion
\cite{Vardi,Jul98,Hop01,Drummond,Mackie,Dam03,HYLing}. The STIRAP
relies on the coherent population trapping (CPT) state, which is a
superposition of two long-lived levels which decouple from the
excited levels \cite{STIRAP1,STIRAP2,STIRAP3}. In the atom-molecule
BEC, the CPT state (also called as the dark state) is a
superposition of the atomic and molecular ground states, and has
been observed in recent experiments \cite{AMDS,PDLett}. In
principle, one can realize a complete and reversible atom-molecule
conversion as long as the system is kept in the CPT state. However,
it was shown the mean-field collisions set a limit to conversion
efficiency by about $46\%$\cite{Drummond,Mackie}.

In Ref. \cite{HYLing}, Ling et al. recognized the CPT condition in the linear atomic system differs from those in the atom-molecule BEC. They
found that high conversion efficiency, namely about $83\%$, is achievable provided that the CPT condition is satisfied during whole time
evolution. In their work a single time-varying laser pulse is used for the bound-bound transition, while a constant Feshbach magnetic field is
applied in the free-bound transition. The Feshbach resonance method was thought to present relatively large free-bound coupling than that of the
photoassociation. However, both methods are actually equivalent in theoretical point of view. In addition, we note that the highest-conversion
efficiency in their work appears near the exact Feshbach resonant, which unfortunately leads to strong atomic loss.

In this paper, we study the STIRAP in the atom-molecule BEC by using the frequency modulation scheme \cite{HYLing}, where two time-varying
Guassian laser pulses are employed. Our scheme has its advantage to provide additional parameters to optimize the conversion efficiency so that
an extreme high conversion efficiency, namely about $92\%$, can be obtained. Moreover, the highest conversion occurs at far off-detuned region
for the free-bound transition, which avoids the loss of atomic BEC in the Feshbach resonance version of the STIRAP. Our paper is organized as
follows. In Sec. II, we present a simple three-level model for the atom-molecule BEC system, and investigate the effects of the mean-field
collisions on the CPT condition and dynamical instability. In Sec. III, we present our numerical results and some discussions. Finally, a
summary of our paper is presented.

\section{Theoretical model}

We consider a $\Lambda$-type atom-molecule system, as shown in Fig. 1 (a), where a pair of free colliding atoms in level $\vert a\rangle$ is
converted to a bound molecule in level $\vert g\rangle$ via an intermediate molecular level $\vert b\rangle$. The free-bound transition $\vert
a\rangle\rightarrow \vert b\rangle $ is characterized by Rabi frequency $\Omega_1$, while the coupling strength $\Omega_2$ corresponds to the
bound-bound transition $\vert b\rangle\rightarrow \vert g\rangle$. In practice, the atoms in level $\vert a\rangle$ is a BEC gas with density
denoted as $\rho$. Due to the Bose enhancement, the free-bound Rabi frequency $\Omega_1\propto \sqrt{\rho}$ \cite{Drummond}. The detuning for
the free-bound and bound-bound transitions are denoted as $\Delta_1=\omega_b-2\omega_a-v_1$ and $\Delta_2=\omega_b-\omega_g-v_2$, respectively,
with $v_\sigma$ the central frequencies of the Laser fields $\sigma=1$, $2$. Thus the two-photon (Raman) detuning is
$\delta=\Delta_1-\Delta_2\equiv(\omega_g-2\omega_a)-(v_1-v_2)$.

Following Refs. \cite{Vardi,Jul98,Hop01,Drummond,Mackie,Dam03,HYLing}, we assume that the atoms and the molecules can be described solely by
three bosonic fields. For an atom-molecule BEC system with high density, we employ further the standard mean-field treatments, i.e., replacing
the field operators by the normalized field amplitudes $a$, $b$, and $g$. From Heisenberg equation of the field operators, we obtain the
following coupled equations:
\begin{mathletters}
\begin{eqnarray}
i\dot{a} &=&(\Lambda _{aa}|a|^{2}+\Lambda _{ag}|g|^{2})a-\Omega
_{1}a^{\ast}b, \\
i\dot{b} &=&\left(\Delta_1-i\gamma _{b}/2\right) b-\frac{1}{2}%
(\Omega _{1}a^{2}+\Omega _{2}g), \\
i\dot{g} &=&(\Lambda _{ag}|a|^{2}+\Lambda _{gg}|g|^{2})g +\delta
g-\frac{1}{2}\Omega _{2}b, \label{equation of motion}
\end{eqnarray}%
\end{mathletters}
where $\Lambda_{ij}=\rho U_{ij}$ represent two-body collision rate with $U_{ii}=4\pi\hbar a_i/m_i$ and $U_{ij}=U_{ji}=2\pi\hbar a_{ij}/\mu_{ij}$
for $i\neq j$. Here $a_{ij}$ are s-wave scattering lengths, and $\mu_{ij}$ are the reduced masses between states $i$ and $j$. We neglect
mean-field interactions related to the excited molecular state due to small occupation of this state \cite{Drummond}. The decay rate $\gamma_b$
is added phenomenologically to take into account the loss of the excited molecules.


\begin{figure}[htp]
\begin{center}
\includegraphics[width=8cm]{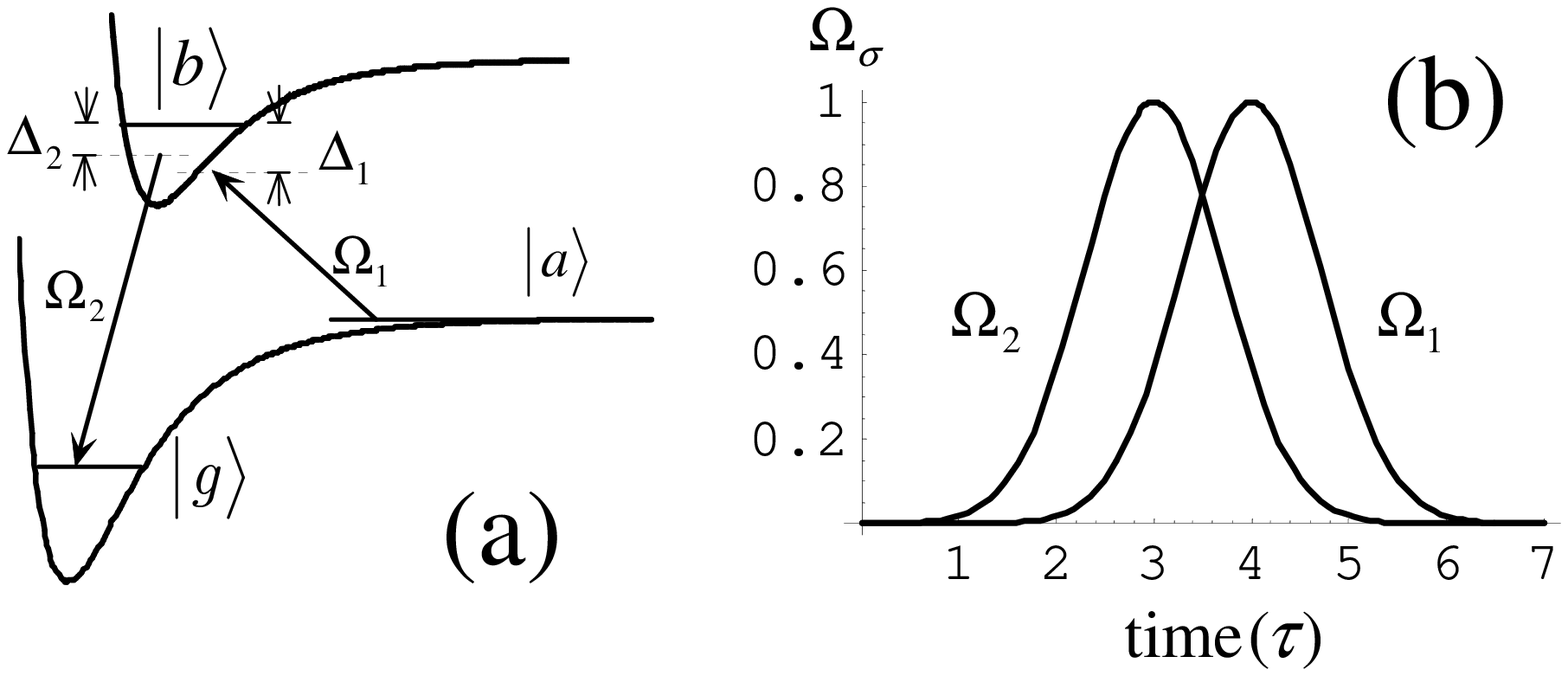}
\vskip -8cm \caption{ (a) Schematic picture of three-level
atom-molecule system; (b) Rabi frequencies as a function of time,
generated by counterintuitive order of the two laser pulses.}
\end{center}
\end{figure}

It was shown that in a linear $\Lambda$ three-level system, the CPT state occurs
if the two-photon resonance condition $\delta=0$ is satisfied
\cite{STIRAP1,STIRAP2,STIRAP3}. However, the mean-field collisions probably modify the CPT condition \cite{HYLing}. To obtain the new condition,
we neglect temporarily the particle losses, which implies the conservation of particle number i.e. $|a|^2+|b|^2+|g|^2=1$. By introducing
$\alpha=\alpha^{(0)}e^{-i\mu_{\alpha}t}$ (for $\alpha=a,b,g$) with the chemical potentials $\mu_{b}=\mu_{g}\equiv2\mu_a$, we can obtain the
stable solution of Eq. (\ref{equation of motion}). The CPT steady state corresponds to $b^{(0)}=0$, and
\begin{eqnarray}
|a^{(0)}|^{2} &=&1-2|g^{(0)}|^{2}  \nonumber \\
&=&\frac{2}{1+\sqrt{1+8(\Omega _{1}/\Omega _{2})^{2}}},\label{CPT}
\end{eqnarray}
which depends only explicitly on the two coupling strengths. For
$\Omega _{1}/\Omega _{2}=0$, $|a^{(0)}|^{2}=1$; for
$\Omega_{1}/\Omega _{2}\rightarrow\infty$, $|g^{(0)}|^{2}=1/2$.
The dotted lines of Fig. 3 present the CPT solutions
$|a^{(0)}|^{2}$ and $|g^{(0)}|^{2}$ for two time-varying laser
fields (see Eq. (\ref{laser pulses}). The CPT solution appears
under generalized two-photon resonance condition \cite{HYLing}:
\begin{equation}
\delta =(2\Lambda _{aa}-\Lambda _{ag})|a^{(0)}|^{2}+(2\Lambda
_{ag}-\Lambda _{gg})|g^{(0)}|^{2}.\label{Generalized CPT
condition}
\end{equation}
In the absence of collisions $\Lambda _{ij}=0$, Eq. (\ref{Generalized CPT condition}) is just the ordinary two-photon resonance condition. For
two Gaussian laser pulses in a counterintuitive order, $\delta$ varies slowly as shown in the inset of Fig. 3.

The mean-field collisions result in dynamical instability of the CPT solutions. To investigate it, we employ the standard perturbation theory,
i.e., adding small perturbations to the CPT solutions as $\alpha=(\alpha^{(0)}+\delta\alpha)e^{-i\mu_{\alpha}t}$, where $\delta
\alpha=\eta_{\alpha} e^{-i\omega t}+\nu_{\alpha}^*e^{i\omega t}$ (for $\alpha=a,b,g$), and $\omega$ is eigenfrequency of the perturbations. The
dynamical instability of the CPT solution takes place when the eigenfrequency is imaginary. In this case, a weak perturbation may induce an
exponential growth of the excited molecules, which in turn lowers remarkably the fraction of ground-state molecules. After some tedious
calculations, one can obtain the eigenfrequency of the perturbations.

\begin{figure}[htbp]
\begin{center}
\begin{picture}(240,190)(0,0)
\put(0,0){\includegraphics[width=8cm]{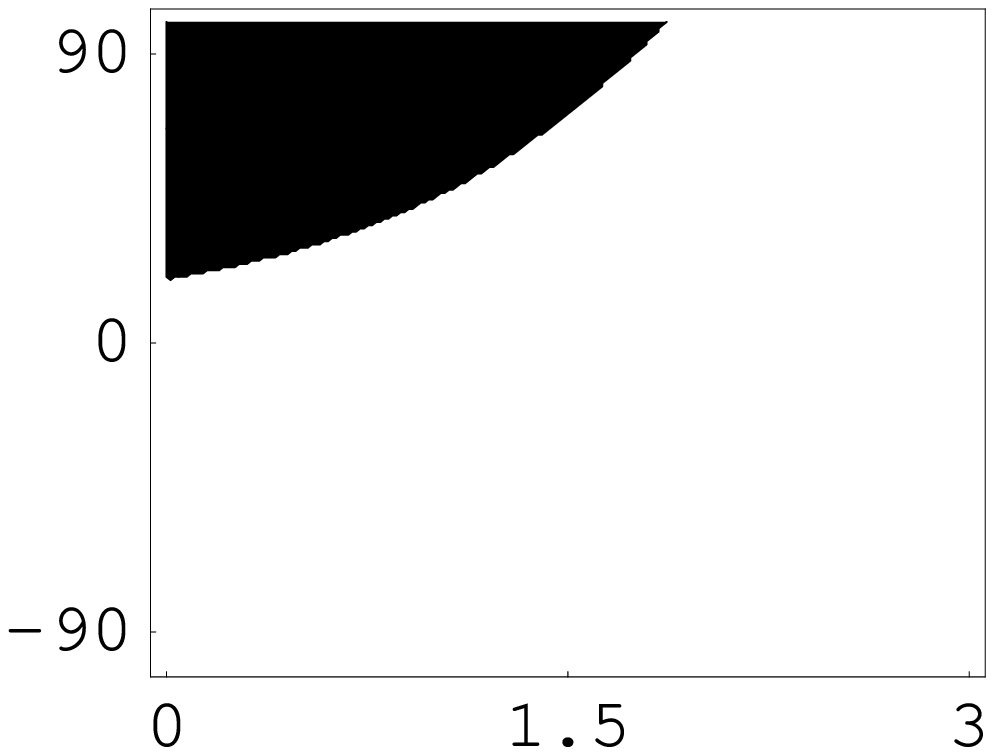}}
\put(70,30){\includegraphics[width=5cm]{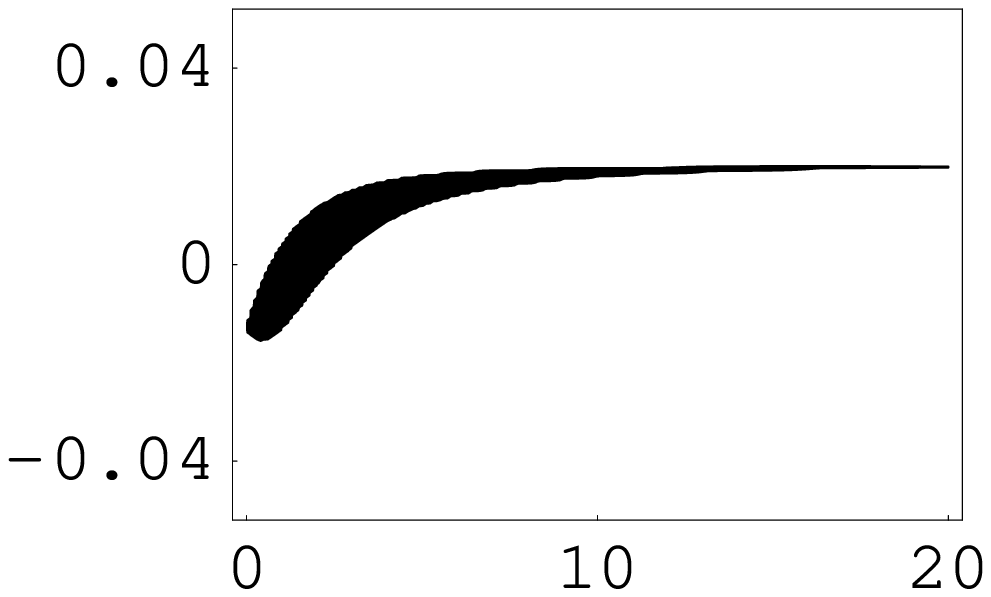}}
\put(115,-10){\makebox(1,1)[l]{{\large $\Omega_2/\Omega_1$}}}
\put(0,100){\makebox(1,1)[l]{{\Large $\frac{\Delta_1}{\Omega_1}$}}}
\put(140,140){\makebox(1,1)[l]{{\Large I}}}
\put(140,90){\makebox(1,1)[l]{{\large II}}}
\end{picture}
\vskip 0.7cm \caption{The dynamically unstable regions in the
($\Omega_2/\Omega_1$, $\Delta_1/\Omega_1$) plane. Other parameters
are taken as (in unit of $\Omega_1=2.1$ MHz):
$\Lambda_{aa}=1.02\times10^{-2}$, $\Lambda_{ag}=-1.32\times10^{-2}$
and $\Lambda_{gg}=0.51\times10^{-2}$. Inset: the unstable region II
near $\Delta_1=0$.}
\end{center}
\end{figure}

In Fig. 2, the instability regions calculated numerically are shown in the ($\Delta_1$, $\Omega_2$) plane, where both $\Delta_1$ and $\Omega_2$
are in unit of $\Omega_1$. Similar to the previous work \cite{HYLing}, there are two unstable areas: the area I takes place only in positive
$\Delta_1$ (i.e., red-detuned region) for small $\Omega_2/\Omega_1$; while unstable region II is a thin area centered at $\Delta_1=\Lambda_{ag}$
for small $\Omega_2/\Omega_1$. Unlike the previous work \cite{HYLing}, part of area II in the small $\Omega_2/\Omega_1$ limit appears in the
blue-detuned region due to negative $\Lambda_{ag}$ for $^{87}$Rb atom-molecule system.

\section{Numerical results}

The atom-molecule conversion efficiency is defined as $\eta=2|g(\infty)|^2$, which measures how many atoms in level $|a\rangle$ converted to the
stable molecule state $|g\rangle$. It was shown the mean-field collisions set a limit to conversion efficiency, and the highest conversion
efficiency is thus about $46\%$ \cite{Drummond}. Ling et al. found however that higher efficiency ($83\%$) can be obtained as long as the CPT
condition, Eq. (\ref{Generalized CPT condition}), is satisfied during whole time evolution \cite{HYLing}. Following their scheme, we restudied
the STIRAP in the atom-molecule BEC system. Two time-varying laser pulses are employed in our model, as shown in Fig. 1 (b). The Rabi
frequencies are taken as
\begin{equation}
\Omega_\sigma(t)=\Omega_0\exp[-(t-t_\sigma)^2/\tau^2],\label{laser
pulses}
\end{equation}
with $\sigma=1$, $2$. To fulfil the adiabatic condition, we take $\Omega_0\tau=5\times 10^{3}$. The two Gaussian pulses are designed in
counterintuitive order: the bound-bound laser field centered at $t_2$ is applied earlier than the free-bound laser field centered at $t_1$. The
delay time between two pulses is defined as $T=t_1-t_2$. Compared with previous work \cite{HYLing}, our scheme provides addition parameters
$\Delta_1$ and $T$, which is useful for optimizing the atom-molecule conversion.

Following Refs. \cite{Drummond,Dam03}, we consider a dilute $^{87}$Rb atomic gas. The peak of Rabi frequency takes the form of $\Omega_0=2.1
\sqrt{\rho/\rho_0}$ MHz, where $\rho_0$ is the peak atomic density. We consider high atomic density with $\rho=\rho_0\equiv4.3\times
10^{14}$cm$^{-3}$, so that $\Omega_0=2.1$ MHz and the pulse duration $\tau=5\times10^3/\Omega_0\simeq2.4$ ms. The decay rate of the excited
molecules is chosen as $\gamma_b=74$ MHz. The s-wave scattering strengthes for $^{87}$Rb atoms are taken as \cite{Drummond}: $U_{aa}=4.96\times
10^{-17}$ MHz cm$^3$, $U_{ag}=-6.44\times 10^{-17}$ MHz cm$^3$ and $U_{gg}=2.48\times 10^{-17}$ MHz cm$^3$. Considering the density
$\rho=4.3\times 10^{14}$cm$^{-3}$, we obtain $\Lambda_{aa}=21.328$ kHz=$1.02\times10^{-2}\Omega_0$, $\Lambda_{ag}=-27.692$
kHz=$-1.32\times10^{-2}\Omega_0$, and $\Lambda_{gg}=10.664$ kHz=$0.51\times10^{-2}\Omega_0$, respectively.

\begin{figure}[tph]
\begin{centering}
\includegraphics[width=9cm]{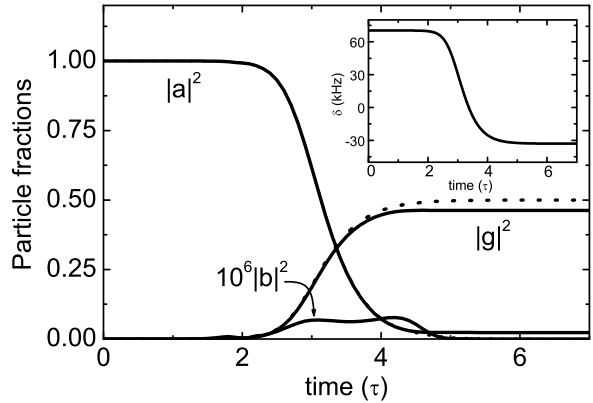}
\end{centering}
\caption{The particle fractions as a function of time (in unit of
$\tau=2.4$ ms). The pulse centers are $t_2=2.5\tau$ and
$t_1=3.77\tau$. The free-bound detuning $\Delta_1=-1.4\gamma_b$. The
dotted lines: the CPT solutions, Eq. (\ref{CPT}). Inset: the
two-photon (Raman) detuning $\delta$ as a function of time $t$, Eq.
(\ref{Generalized CPT condition}).} \label{fig2}
\end{figure}

In Fig. 3, we calculate time evolution of the three-component populations for a fixed free-bound detuning $\Delta_1=-1.4\gamma_b$, where a
blue-detuned laser is adopted to avoid dynamical unstable region. Mean-field collision terms result in dynamical instability of the CPT state
for positive $\Delta_1$, which in turn leads to lower conversion efficiency \cite{HYLing}. As show by the inset of Fig. 3, with time evolution
$\delta$ deceases slightly according to the CPT condition, Eq. (\ref{Generalized CPT condition}). In practice, such laser frequency modulation
can be realized by setting $\Delta_2=-1.4951\gamma_b$ initially, and then increasing it adiabatically to $-1.3554\gamma_b$. Numerical results of
$|\alpha|^2$ ($\alpha=a, b, g$) are shown by solid curves, which are consistent with the the CPT solutions $|a^{(0)}|^2$ and $|g^{(0)}|^2$. We
find that the final population of the stable molecules $|g(\infty)|^2=0.46$, which means that $92\%$ atoms in level $|a\rangle$ are converted
into the stable molecules.

We investigate further the relation between the delay time $T$ and $|g(\infty)|^2$ with the free-bound detuning $\Delta_1$, where we fix
$t_2=2.5\tau$ and increase $T$, so that $t_1=3.0\tau$, $3.77\tau$, and $4.5\tau$. As shown by the solid line of Fig. 4, the optimal conversion
occurs at $t_1=3.77\tau$. The molecular population $|g(\infty)|^2$ (also $\eta$) varies slowly with $\Delta_1$ in the blue-detuned region
(negative $\Delta_1$), while it decreases quickly to zero for positive $\Delta_1$. This asymmetric variation is aroused from dynamical
instability of the CPT in the red-detuned region \cite{HYLing}. For the case of $t_1=3\tau$ (dashed line), however, $|g(\infty)|^2$ begins to
decrease in the blue-detuned region since for certain parameters the instability can also occur as shown by inset of Fig. 2. It should be
mentioned that in previous work \cite{HYLing} the highest conversion appears at the exact resonance ($\Delta_1=0$). If the free-bound transition
is realized by using Feshbach resonance, strong atomic losses may take place near the resonant point. In our scheme, however, the highest
conversion occurs at $\Delta_1=-1.4\gamma_b$, far-off resonance to the Feshbach transition. Therefore, condensate atomic losses in state
$|a\rangle$ can be ignored safely.

In the above theoretical treatments, we have neglected the decay of stable molecules, $\gamma_g$. In recent experiment \cite{AMDS}, however, it
was found that $\gamma_g$ is relatively large and depends on the intensity of the first laser, which may lead to very low conversion rate. There
are several reasons that lead to dramatic losses of the ground-state molecules. Firstly, the counter-rotating couplings of the two Raman lasers
give contribution to incoherent loss of the stable molecules. Secondly, the coupling of other vibrational molecular levels near level
$|b\rangle$ break the ideal three-level system, and provide addition loss channels. Finally, the rogue dissociation of the molecule itself also
give negative contributions to the atom-molecule conversion. In fact, the obtained molecular fraction $|g(\infty)|\sim 10^{-4}$, i.e., only
hundreds of ground-state molecules are prepared from the $^{87}$Rb BEC of about $10^5$ atoms\cite{AMDS}. Such a low conversion is caused not
only from the decay of the stable molecules mentioned above, but also because of an extremely weak free-bound coupling used in the experiment.
To overcome the latter problem a large Franck-Condom overlap integral for the free-bound transition is crucially needed. By carefully choosing
suitable molecular levels, one can increase the conversion efficiency by several orders of magnitude.

\begin{figure}[tph]
\begin{centering}
\includegraphics[width=9cm]{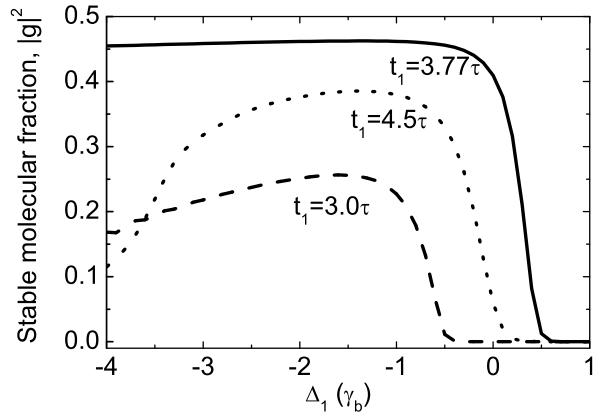}
\end{centering}
\caption{The fraction of the stable molecules as a function of the
detuning $\Delta_1$ for various time delays. The second pulse center
$t_2=2.5\tau$ is fixed. By increasing $T$, the first pulse centers
are chosen as $t_1=3.0\tau$ (the dashed line), $3.77\tau$ (the solid
line), and $4.5\tau$ (the dotted line). }
\end{figure}

\section{Conclusion}

In summary, the efficient conversion between atomic BEC's and the stable molecular BEC's has been investigated by using the frequency modulation
STIRAP scheme. The effects of the mean-field collisions on the CPT condition and dynamical instability are also discussed. We find that
dynamical instability of the CPT solution can also occurs in the blue detuned region due to negative atom-molecule interactions. A very high
conversion efficiency about $92\%$ can be obtained by choosing optimally the free-bound detuning and time delay of the two Raman lasers. We also
find that the highest conversion occurs at the far off-detuned free-bound transition which is important for Feshbach resonance version of the
STIRAP in the atom-molecule BEC.

\section*{acknowledgement}

We thank Profs. C. K. Kim, K. Nahm for helpful discussions. This study was financially supported by Pusan National University in program
Post-Doc 2006 and Korea Research Foundation Grant (KRF-2006-005-J02804).



\begin{thebibliography}{*}

\bibitem[*]{email} Present address: Department of physics, School of
Science, Beijing Jiaotong University, Beijing 100044, China\\
Electronic address: grjin@bjtu.edu.cn

\bibitem[$\dagger$]{email1} Electronic address: swkim0412@pusan.ac.kr

\bibitem{Wynar} R. Wynar {\it et al.}, Science {\bf 287}, 1016 (2000).

\bibitem{drum} P. D. Drummond {\it et al.}, Phys. Rev. Lett. {\bf 81}, 3055
(1998).
\bibitem{Java99} J. Javanainen and M. Mackie, Phys. Rev. A {\bf 59}, R3186
(1999).
\bibitem{hein} D. J. Heinzen {\it et al.}, Phys. Rev. Lett. {\bf 84}, 5029
(2000).

\bibitem{timm} E. Timmermans {\it et al.}, Phys. Rev. Lett. {\bf 83}, 2691
(1999); Phys. Rep. {\bf 315}, 199 (1999).

\bibitem{Donley} E. A. Donley {\it et al.}, Nature {\bf 417}, 529
(2002); N. R. Claussen {\it et al.}, Phys. Rev. A {\bf 67}, 060701
(R) (2003).

\bibitem{Meystre} D. Meiser and P. Meystre, Phys. Rev. Lett. {\bf
94}, 093001 (2005).
\bibitem{GRJIN} Guang-Ri Jin, Chul Koo Kim, and Kyun Nahm, Phys. Rev. A {\bf 72}, 045602 (2005)


\bibitem{Vardi} A. Vardi, D. Abrashkevich, E. Frishman, and M. Shapiro, J. Chem.
Phys. {\bf 107}, 6166 (1997); A. Vardi, V.A. Yurovsky, and J.R.
Anglin, Phys. Rev. A {\bf 64}, 063611 (2001).

\bibitem{Jul98} P.S. Julienne, K. Burnett, Y.B. Band, and W.C. Stwalley,
Phys. Rev. A \textbf{58}, R797 (1998).


\bibitem{Hop01} J.J. Hope, M.K. Olsen, and L.I. Plimak, Phys. Rev. A {\bf 63},
043603 (2001).

\bibitem{Drummond} P. D. Drummond, K.V. Kheruntsyan, D.J. Heinzen, and R.H. Wynar, Phys.
Rev. A {\bf 65}, 063619 (2002); P.D. Drummond, K.V. Kheruntsyan,
D.J. Heinzen, and R.H. Wynar, Phys. Rev. A {\bf 71}, 017602
(2005).

\bibitem{Mackie} J. Javanainen and M. Mackie, Phys. Rev. A {\bf
58}, R789 (1998); M. Mackie, A. Collin, and J. Javanainen, Phys.
Rev. A {\bf 71}, 017601 (2005).

\bibitem{Dam03} B. Damski, L. Santos, E. Tiemann, M. Lewenstein,
S. Kotochigova, P. Julienne, and P. Zoller, Phys. Rev. Lett. {\bf
90}, 110401 (2003).

\bibitem{HYLing} H.Y. Ling, H. Pu, and B. Seaman, Phys. Rev. Lett.
{\bf 93}, 250403 (2004). Phys. Rev. A {\bf 72}, 013608 (2005).

\bibitem{STIRAP1} S. E. Harris, Phys. Today {\bf 50}, 36 (1997).

\bibitem{STIRAP2} M. O. Scully, and M. S. Zubairy, {\it Quantum optics}
(Cambridge University Press, Cambridge, 1997).

\bibitem{STIRAP3} N. V. Vitanov et al., Annu. Rev. Phys. Chem. {\bf 52}, 763
(2001).



\bibitem{AMDS} K. Winkler, G. Thalhammer, M. Theis, H. Ritsch, R.
Grimm, and J. H. Denschlag, Phys. Rev. Lett. {\bf 95}, 063202
(2005).
\bibitem{PDLett}R. Dumke, J. D. Weinstein, M. Johanning, K. M. Jones, and
P. D. Lett, Phys. Rev. A {\bf 72}, 041801 (R) (2005).

\end{thebibliography}
\end{document}